\documentclass[useAMS,usenatbib]{mn2e}
\input psfig.tex

\title[BO effect at $z\sim0.35$]{The Butcher--Oemler effect at $z \sim 0.35$: 
a change in perspective.}
\author[Andreon et al.]{S. Andreon,$^1$\thanks{andreon@brera.mi.astro.it}
H. Quintana,$^2$ M. Tajer,$^1$ G. Galaz,$^2$ J. Surdej,$^3$  \\
$^1$INAF--Osservatorio Astronomico di Brera, Milano, Italy\\
$^2$Departamento de Astronom\'\i a y Astrof\'\i sica, Pontificia Universidad 
Cat\'olica de Chile, Santiago, Chile\\
$^3$Institut d'Astrophysique et de G\'eophysique, Universit\'e de Li\`ege, 
Sart Tilman, Belgium\\
}
\date{Accepted ... Received ...}
\pagerange{\pageref{firstpage}--\pageref{lastpage}}
\pubyear{2004}
\begin{document}
\maketitle

\label{firstpage}

\begin{abstract}

The present paper focuses on the much debated Butcher-Oemler effect:
the increase with redshift of the fraction of blue galaxies in
clusters. Considering a representative cluster sample made of seven
group/clusters at $z \sim 0.35$, we have measured
the blue fraction from the cluster core to the cluster outskirts and
the field  mainly using wide field CTIO images. This sample represents a random
selection of a volume complete x-ray selected cluster sample, selected so that there
is no physical connection with the studied quantity (blue fraction),
to minimize observational biases. In order to statistically
assess the significance of the Butcher--Oemler effect, we introduce
the tools of Bayesian inference.
Furthermore, we modified the blue fraction definition in order to take into
account the reduced age of the universe at higher redshifts, because
we should no longer attempt to reject an unphysical 
universe in which the age of the Universe does depend on redshift,
whereas the age of its content does not. 
We measured the blue fraction from the cluster center
to the field and we find
that the cluster affects the properties of the galaxies up to two
virial radii at $z\sim0.35$. Data suggest 
that during the last 3 Gyrs no evolution
of the blue fraction, from the cluster core to the field value, is
seen beyond the one needed to account for the varying age 
with redshift of the
Universe and of its content. The agreement of the radial profiles of
the blue fraction at $z=0$ and $z\sim0.35$ implies that the pattern
infall did not change over the last 3 Gyr, or, at least, its variation
has no observational  effect on the studied quantity. 
\end{abstract}


\section{Introduction}

The nature and the time scale of the processes that shape galaxy properties
in clusters and groups are still unclear. The presence of a hot intercluster
gas observed in X-rays should have a role in shaping some galaxy properties
(e.g. Gunn \& Gott 1972). The window opened by the redshift dependence of the
galaxy properties has been used to set constraints on the time scales of
the processes (e.g. Butcher \& Oemler 1978, 1984; Dressler et al. 1997;
Stanford, Eisenhardt \&  Dickinson, 1998; Treu et al. 2003). However, the
observational evidence of the environmental effect is still uncertain. For
example, the existence of a Butcher--Oemler (BO) effect (Butcher \& Oemler 1984),
i.e. the fact that clusters at higher redshift have a larger fraction of
blue galaxies, $f_b$, is still controversial. The controversy is raised by
two criticisms concerning measurements and sample.

Andreon, Lobo \& Iovino (2004; hereafter ALI04) analyse three clusters at
$z\sim0.7$ without finding evidence of a high blue fraction with respect to
$z\sim0$. They also show the drawbacks of the various definitions of $f_b$
adopted in the literature. They conclude then that ``twenty years
after the original intuition by Butcher \& Oemler, we are still in the process
of ascertaining the reality of the effect". The same work put in a different
perspective the results of Rakos \& Shombert  (1995), clarifying the fact
that even if
all the galaxies in the Universe are passively evolving, the blue fraction will
be $f_b\approx 1$ at $z\ga0.7$ in the Rakos \& Shombert (1995) scale.
Therefore, the very high fraction they found at high redshift does not require
any special mechanism to account for the present day counterparts other than
ageing. ALI04 introduce also a first discussion about the difficult task of
measuring the error on $f_b$, given the observations, showing that at least
some previous works have underestimated errors and, by consequence, overstated
the evidence for the BO effect. The role of the inference, the logical step
going from the observed data to the true value and its error, has been further
elaborated in D'Agostini
(2004) in the more general case of unknown individual membership for the
galaxies.

Kron (1994) claimed that all the ``high" redshift clusters known in
the early  80's  ($z\approx 0.3-0.5$) were somewhat extreme in their
properties, and this was precisely the reason why they were detected.
Andreon \& Ettori (1999) quantify this issue, and show that many of
the clusters compared at different redshifts have also different
masses (or X-ray luminosities), in such a way that ``we are comparing
unripe apples with ripe oranges in understanding how fruit ripens"
(Andreon \& Ettori 1999). Together with  Allison-Smith et al. (1993)
and Andreon \& Ettori (1999), ALI04 show that the optical selection of
clusters is prone to produce a biased - hence inadequate - sample for
studies on evolution since at larger redshifts it naturally favours
the inclusion in the sample of clusters with a significant blue
fraction. They show that clusters with a blue fraction as the observed
ones are over-represented in optical cluster catalogs by a factor two,
with respect to identical clusters but without a bursting population.

There is therefore a compelling need to study the properties of galaxies in
clusters at intermediate redshift ($z\approx0.35$), avoiding the bias of an
optical selection, by choosing clusters of the same mass as present day 
studied clusters to avoid an ``apple vs orange" issue. This is the aim of this
paper, where we present a BO--style study of a small but representative sample
of 7 clusters, X-ray selected, of low to average mass (velocity dispersion) and
at intermediate redshift.

The layout of the paper is the following. In sect. 2 we present
optical imaging  and spectral data. In sect. 3 we show that the
studied sample is both representative and X--ray selected. We
revisit in section
4 the definition of the blue  fraction, in order to account
for the reduced age of the universe at higher redshift. Sect. 5
presents some technical details. Results are summarized in Sect. 6,
whereas Sect. 7 discusses relevant results published in the literature
and some final conclusions. Appendixes present a Bayesian
estimate of cluster velocity dispersion, richness, and blue fraction. 

We adopt $\Omega_\Lambda=0.7$, $\Omega_m=0.3$ and $H_0=70$ km s$^{-1}$
Mpc$^{-1}$.

\begin{table}
\caption{The cluster sample}
\begin{tabular}{l r @{.} l r r r r r}
\hline
 Name  
& \multicolumn{2}{c}{z} &
$N_{z}$ & $\sigma_v$ & error & $r_{200}$ \\
\multicolumn{2}{c}{} & & & km/s & km/s & Mpc \\
\hline
XLSSC 024  &  
0&29  & 11 &  430 & 96 & 1.0 \\  %& $3.1 \pm 0.7$ \\
XLSSC 028  &  
0&30  &  8 &  376 & 98 & 0.8 \\ %& $5.3 \pm 1.4$ & \\
XLSSC 009  &  
0&33  & 12 &  236 & 52 & 0.5 \\ %& $6.4 \pm 2.0$ & \\
XLSSC 010  &  
0&33  & 11 &  367 & 96 & 0.8 \\ %& $17.2 \pm 3.2$ & \\
XLSSC 016  &  
0&33  &  5 &  915 & 294 & 2.0 \\ %& \\
XLSSC 006  &  
0&43  & 39 &  837 & 106 & 1.7 \\ %& $163 \pm 22$ & \\
XLSSC 012  &  
0&43  & 12 &  741 & 165 & 1.5 \\ %& $18.1 \pm 4.0$ & \\
\hline                                                   
\end{tabular} \hfill \break
{ }
\end{table}

\section{The data \& data reduction}

\subsection{Photometry}

We use the same imaging data as Andreon et al. (2004a), with some 
additional observations taken in 2002 with the same
instrument and telescope. Briefly, optical $R$-- and $z'$--band
($\lambda_c\sim9000${\AA}) images were obtained at the Cerro Tololo
Inter--American Observatory (CTIO) 4m Blanco telescope during three
observing runs, in August 2000, November 2001 and September 2002
with the Mosaic II camera. Mosaic II is a 8k$\times$8k camera with a $36
\times 36$ arcminute field of view. Typical exposure times were 1200
seconds in $R$ and $2 \times 750$ seconds in $z'$. Seeing in the final
images was between 1.0 and 1.4 arcseconds Full--Width at Half--Maximum
(FWHM) in the September 2002 and November 2001 runs, and 0.9 to 1.0
arcsec FWHM during the August 2000 run. The useful nights of the three
observing runs were photometric. Data have been reduced in the standard
way (see Andreon et al. 2004a for details). 

Source detection and characterization were performed employing SExtractor v2
(Bertin \& Arnouts 1996). Colours and magnitudes are computed within a fixed 5
arcsecond radius aperture. A larger aperture, for colours, is used here with
respect to  Andreon et al. (2004a), where 1.9 arcsec was used, in order not to 
miss any potential  star formation occurring at radii not sampled by the
previously adopted aperture. Of course, results of that paper are unaffected by
our present aperture choice. 

Object magnitudes are quoted in the photometric system of the
associated standard stars: $R$ magnitudes are calibrated with
Landolt (1992) stars, while $z'$ magnitudes are calibrated with SDSS
(Smith et al. 2002) standard stars.  We keep $R$ and $z'$ magnitudes in
their system (i.e. Vega and SDSS, respectively).

\subsection{Spectroscopy}

Our clusters have been observed spectroscopically at Magellan, NTT or
VLT (see Willis et al. 2005). Redshifts for a
minimum of 5 up to 39 cluster members have been acquired per cluster
with typical individual  errors on redshift of 50 to 150 km/s
(depending on instrument, exposure time, etc.), as detailed in the
mentioned papers.

Velocity dispersions are computed using the Beers et al. (1990)
scale estimator, as detailed in the Appendix and are listed together
with their errors in Table 1.

\section{The cluster ``Apple vs orange" issue}

As discussed in the Introduction, the cluster selection criteria
should not bias the targeted measurement (the blue fraction). As
mentioned, the optical selection, especially if performed in the blue
band rest-frame, boosts by  construction the blue fraction at high
redshift, unless some precautions are taken. The X--ray selection is useful
because the cluster X--ray emissivity is not physically related, in a
cause--effect relationship, to the colour of cluster galaxies, the
other factors (e.g. mass, dynamical status, etc.) being kept fixed. 
Fairley et al. (2002) and Wake et al. (2005) exploit a similar X--ray
selection,  for a cluster sample much more (the formers) or slightly
more (the latters) massive (X--ray bright), but statistically
uncontrolled.

The cluster sample studied in this paper is not an uncontrolled collection
of clusters, but a random sampling of an X--ray flux limited sample of
clusters in a narrow redshift range ($0.29\la z<0.44$), drawn from  the
ongoing XMM-LSS survey (Pierre et al. 2004, and Pierre et al. in
preparation). The clusters actually used in the present paper are listed
in  Andreon et al. (2004a), or presented in a future catalogue. The
sample studied here is a purely X--ray selected one drawn from a sample
constructed using both an X--ray and optical selection criteria (the
XMM-LSS survey),  as clarified below. We refer to Pierre et al. (2004) for
details about the XMM-LSS survey, and we discuss here only some relevant
points.

One great advantage of a volume complete sample (or a random sampling of a
volume complete sample) over an uncontrolled one is that each object has a
chance of occurring that is proportional to its number density, i.e. occurs
in the sample with the same natural frequency it occurs in the Universe. The
above property is especially useful when computing ensemble averages (like
composite clusters), because it makes the statistical analysis
straightforward. Instead, averages performed over uncontrolled samples
(e.g. combined ``clusters"  formed by staking clusters from uncontrolled
lists) lack predictive power because the sample representativity is
unknown.  An astronomical example, together with a real-life application
of  the above concept is discussed in Sect 6.3.1.

\subsection{Malmquist-like (or Eddington-like) biases on $f_b$: redshift range 
selection}

The precise choice of a redshift range largely depends on the quality of
the available optical photometry and on the availability of velocity
dispersions. The lower redshift limit ($z\sim0.29$) has been chosen because
of saturation issues: our images are exposed too long for brighter
objects and their cores saturate, because exposure time has  been
originally optimized for the detection of $z\sim1$ galaxies. The fuzziness
of the nearest redshift limit is due to varying seeing conditions and sky
brightness during the observing runs.

The upper redshift limit ($z=0.44$) comes from  our desire to get a
complete and unbiased sample. At $z>0.44$ not all clusters have a known
velocity dispersion, and it is legitimate to suspect that clusters without
a known $\sigma_v$ have a different blue fraction from clusters with a
known $\sigma_v$, all the remaining parameters being kept fixed. Indeed,
a cluster with a larger number of red galaxies has, observationally,
better chance of having a larger number of confirmed members than an equally
rich, but poor in red galaxies, cluster, because background galaxies
are more aboundant among blue galaxies in percentage.
Clusters rich in blue galaxies may have so few
confirmed members that a cluster velocity dispersion cannot be computed
with a sufficient accuracy. Therefore, a cluster with a small blue fraction
has a better chance to have a measured velocity dispersion than one with a 
large blue fraction. Below $z=0.44$, all clusters have a known
velocity dispersion and this problem does not arise.

In general, an upper redshift limit is needed for another reason:  we want
the faintest considered galaxies to be still affected by a negligible
photometric error (see below), because it is quite dangerous to attempt to
correct the blue fraction for the bias induced by photometric errors.  In
fact, the uneven colour distribution of galaxies (for example $f_b=0.2$
means that more than 80 \% of the galaxies have colours in a narrow red
color range, and the remaining 20 \% are spread over a large blue color
range) and errors on colours of 0.2 mag amplitude  produce a large
Malmquist-like (or Eddington-like) bias, difficult to correct for without
knowing the galaxy colour distribution, as first explained by Jeffreys
(1938).  The Eddington (1940) reply to the Jeffreys (1938) paper clarifies
that improved values, i.e. corrected by the error measurements, ``should
not be used for any kind of statistical inquiry" in good agreement with
Jeffreys (1938). If the ultimate limit of the measurements lays in
photometric errors, it is perhaps preferable to increase the quality of the
photometry, rather than increasing the size of the sample, and therefore we
prefer to have a small, but high quality sample, than a large, low quality
one. 

Malmquist-like
biases affect our blue fraction determination at $z>0.44$, and therefore
are of no concern for our analysis. However, they may be a concern for
other similar works. The above Malmquist-like bias, joined to the use 
of data with a fixed quality (such as those
coming from surveys) both unduly increase
the observed fraction of blue galaxies
with redshift, simply because galaxies become fainter and
photometric errors increase with redshift. The above effect has
nothing to do with the Butcher-Oemler effect, of course, because
the amplitude of the effect depends on the data quality,
not on the galaxy properties.

\subsection{Which selection criteria?}

The sample from which we have drawn our clusters is formed by all
clusters detected {\it both} in X--rays and in the colour space. 
Details about the colour detection can be found in Andreon et al.
(2004a,b). At the redshift studied in this paper, clusters stand out
in the colour--space, and also in the direct--space (i.e. in the sky plane) as
shown in section 3.2 of Andreon et al. 2004a, i.e. the probability to
miss in the optical a cluster in the  considered redshift range is
virtually zero. In particular, clusters at  $z\leq 0.29$ stand out in
the direct--space (i.e. on images) so conspicuously that their brightest
galaxies saturate the instrument (exposure time is tuned for
$z\sim1$ galaxies). Can a cluster get unnoticed when its galaxies
(almost) saturate the instrument? Therefore, even if in principle our
cluster sample is drawn from a sample that uses two criteria (X--ray
emission and colour-detection), at the studied redshifts the colour
selection does not bias the cluster selection because it does not
filter out any object. To check the above, during the spectroscopic
campaign we  devoted a (small) fraction of time to
spectroscopically confirm candidates not meeting the colour detection.
None turns out to be confirmed in the considered redshift range,
showing that if clusters of galaxies not detectable in the colour-space do
exist, they are so rare that they are not likely to occur in a sample
like ours. As an independent check, we spectroscopically confirmed
colour detected clusters without detectable X-ray emission, at the
same and higher redshift, showing that the optical selection goes
deeper in the cluster mass function than the X-ray selection.  One
such an example, RzCS 001 at $z=0.49$ is listed in Andreon et al.
(2004a). Another one, RzCS 052 at $z=1.02$ is studied in Andreon et
al. (2005).  The presence of other clusters deliberately
not studied in this paper
in the very same studied volume of Universe, such as RzCS 001, 
emphasizes once more that we are studing an {\it x-ray selected} 
cluster sample and clarifies that the adopted selection is
a deliberated choice in order to avoid the bias of the optical
selection at high redshift, not an obliged choice dictated by our 
ignorance about which other clusters are present 
in the studied volume of the Universe.

\subsection{Random sampling from a complete sample}

Inside the selected redshift range, we removed all clusters with
$r_{200}$ radii overlapping each other {\it in the sky plane} or
which exceed the studied field of  view of each individual CTIO
pointing ($\sim 0.3$ deg$^2$ area, to keep uniform quality all across
the area), as well as one XLSSC cluster that lacks an obvious center.
These (observational--driven) cluster selections are independent on
the cluster blue fraction and hence produce no biases. Therefore, our
sample constitutes a random sampling of XMM-LSS clusters in the selected
redshift range.

\subsection{Details about the X-ray selection}

As mentioned, our sample is drawn from the XMM-LSS, and therefore
our sample inherits its advantages and limitations.
To a first approximation, the survey is flux limited, and 
therefore brighter 
clusters, visible over larger volumes are in principle over-represented
in the survey. However, here the studied redshift 
interval is small ($\Delta z = 0.14$), and the effect should be minor. 

Furthermore, the XMM-LSS is surface brightness
limited too, as most existing surveys, in spite of the use of
wavelets in the detection step to mitigate surface brightness effects.
Extensive numerical simulations (Pacaud et al., in preparation)
show that, for core radii typical of the studied objects,
detectability is larger than 90 \% for all our objects.

X-ray fluxes inside half the optical $r_{200}$ radius (computed as
specified in Sec 5) were computed in the $0.5 - 2$ keV band from MOS1, MOS2
and pn  merged images processed as in Chiappetti et al. (2005). We assumed
a Raymond - Smith spectrum with $kT = 2$ keV and  $z=0.35$, and the average
galactic column density in the XMM-LSS  (Dickey \& Lockman, 1990).
We found four our systems  values in the range 
$0.3 \la Lx \la  16 \ 10^{43}$ erg s$^{-1}$ cm$^{-2}$ in the $0.5 - 2$ keV band.

\medskip

To summarize, the studied sample has $0.3 \la Lx \la  16 \ 10^{43}$,
and it has been selected in a redshift-luminosity-surface brightness
region where detectability is near 100 \%, so that each cluster has the same probability of
occurring in our sample as in the Universe.

\begin{figure*}
\psfig{figure=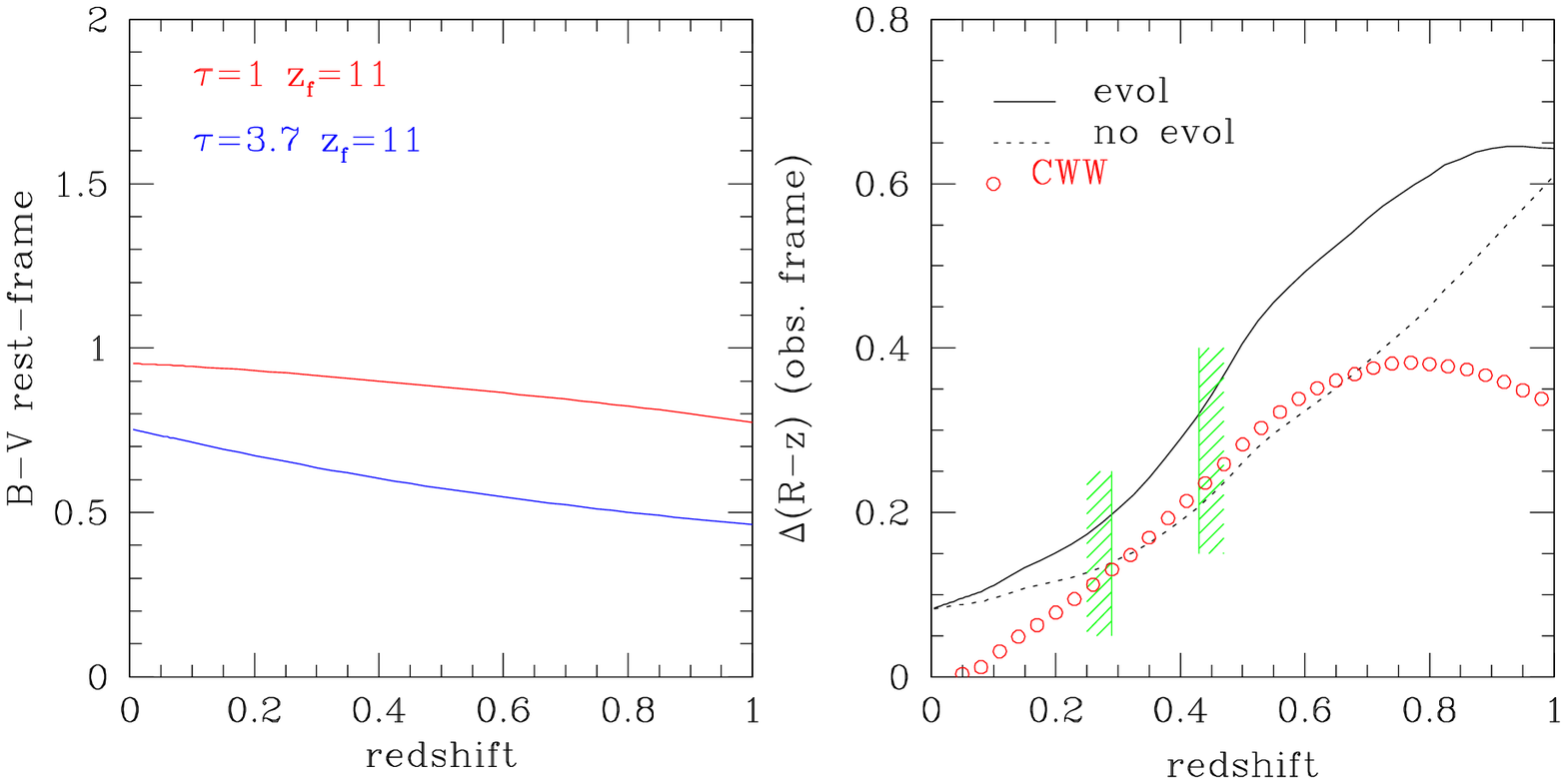,width=15truecm,clip=}
\caption[h]{{\it Left panel:} rest--frame $B-V$ colour of a $\tau=1$ Gyr and
$z_f=11$ stellar population (top red line, mimicking an E), and a 
stellar population having the same $z_f$ but being 0.2 mag bluer
today, i.e. $\tau=3.7$ Gyr, $z_f=11$, referred in this paper as to Sa
(bottom blue line). {\it Right panel}: $R-z'$ colour difference in the
observer rest
frame between the two above stellar populations (solid line). The
dotted line represents the difference for the case of non aging stellar
populations. Dotted circles
are derived using the Coleman, Wu and Weedman (non--evolving) templates. The two  
vertical (green) delimiters mark the redshift range probed in this paper.
}
\end{figure*}

\section{The galaxy ``Apples vs oranges" issue}

Butcher \& Oemler (1985) define the fraction of blue galaxies in the
cluster, $f_b$, as being the fraction of galaxies bluer, by at least
$\Delta=0.2$ mag in the $B-V$ rest--frame, than early--type galaxies at the
cluster redshift (the cluster red sequence). The galaxies have to be
counted down to a given absolute magnitude which is chosen to be
$M_V=-19.3$ mag in our cosmology ($-20$ mag in BO cosmology), within a
reference radius  that encompasses a given fraction of the cluster. 
Moreover, galaxies located in the background or foreground of the cluster
must be removed, for example by statistical subtraction.

The {\it actual} limiting magnitude used in the BO paper is, at the BO 
high redshift end, brighter than $M_V=-19.3$ mag in our cosmology 
($-20$ mag in the BO cosmology) as shown by de Propris et
al. (2003), i.e. different from what the BO definition requires.  A
brighter limiting magnitude at higher redshift is the correct choice if
one wants to track the same population of galaxies at different
redshifts, because of average luminosity evolution experienced by
galaxies. Galaxies having at $z=1$ $M_V=-19.3$ mag are now (at $z=0$)
much fainter than  the $M_V=-19.3$ mag cut. A fixed magnitude cut
therefore does not select similar galaxies at different redshifts,
whereas an evolving limit does. Therefore, we have adopted an evolving
mag limit, as  {\it actually} adopted by BO themselves. 
An evolving limiting magnitude has also been adopted by de
Propris et al. (2003), Ellingson et al. (2001) and ALI04 in their
BO--style studies.

ALI04 discuss the large impact that apparently
minor differences on the $f_b$ definition have on the observed
$f_b$. They found that:

-- the reference colour of the early--type galaxies to be used is the
observed colour of the red sequence, and not the colour of a present day
elliptical, unless we are happy with an evolving $f_b$ fraction for a
sample of galaxies passively evolving;

-- the reference radius should scale with the cluster size, and not
be a fixed metric radius, potentially sampling the center of rich
and large clusters and the whole cluster for small groups (another
``apples vs oranges" issue);

-- a unique $\Delta$ should be taken (equal to 0.2 in the $B-V$
rest--frame). If different values are chosen at different redshifts, 
it becomes difficult to
compare populations selected with heterogeneous choices. 

Let us discuss, and revise, the $\Delta$ choice.

There is little doubt that galaxies at higher redshift have younger
stars than present day galaxies, as measured by the fact that the
reddest galaxies have a colour that becomes bluer in the rest--frame
with increasing redshift (e.g. Stanford et al. 1998, Kodama et al.
1998,  Andreon et al. 2004a). This is also the natural outcome of the
current cosmological model that allocates a shorter
age of the universe at higher redshifts. 
At the time of the BO paper, the measurement 
of the blue fraction was a valuable evidence to rule out a 
non-evolving universe. However,
if the aim of deriving the $f_b$ fraction is to measure
an evolution beyond the one due to
the younger age of the universe at high redshift, 
we propose a different choice for $\Delta$, using an
evolving spectral template in order to coherently separate
blue galaxies from red ones. 
This is also an observationally obliged choice, as shown below.

Figure 1 clearly illustrates for our choice. The left panel shows the rest--frame $B-V$
colour of $\tau=1$ (upper curve) and  $\tau=3.7$ (lower curve)
Bruzual \& Charlot (2003) stellar populations of solar metallicity
for exponentially declining star formation rate models, where $\tau$
is the $e$--folding time in Gyr. The formation redshift,  $z_f=11$,
and $e$--folding time, $\tau=1$, are both chosen to reproduce the
observed $R-z'$ colour of our clusters over $0.3 \la z < 1$ (those of
this paper, and those presented in Andreon et al. 2004a), and the
typical colour of present--day ellipticals, $B-V\sim0.95$ mag.  This
population is referred as to the spectro--photometric elliptical one.  
The $e$--folding time of the bluer track is chosen to have a present day
colour $B-V=0.75$ mag, i.e. 0.2 mag bluer than an elliptical, as the BO
definition requires (i.e. $\Delta=0.2$ mag). We refer this template as to
the spectro-photometric Sa, for sake of clarity.  In agreement with
Butcher--Oemler, at $z\sim0$  this spectral template is the
appropriate one to discriminate between red and blue galaxies. However,
the two tracks do not run parallel, which means that
what is characterized today by $\Delta=0.2$ mag was $\Delta>0.2$ mag 
in the past (at higher redshift). This reflects the fact that at 
that time the universe,
and its content, were younger. The choice of a fixed $\Delta$ allows 
galaxies, even those with simple exponential
declining star formation rates, to move from the blue to the red
class, as time goes on (as redshift becomes smaller). That drift boosts  
the blue fraction $f_b$ at high redshift. 
Since the
choice of a fixed $\Delta$ allows a possible drift from one class to the
other, and assuming that a redshift dependence is found for the blue
fraction, does the above 
tell us something about the relative evolution of red and blue 
galaxies? It merely reflects a selection bias related to the way
galaxies are divided in colour classes: a 
class naturally gets contaminated by the
other one. This is precisely what Weiner et al. (2005) observed.

From an observational point of view, measurements are rarely taken in filters
that perfectly match $B$ and $V$. Therefore, the colour cut is computed using a
spectral template. The latter is usually taken from the Coleman, Wu \& Weedman
(1980) spectrum, i.e. for a non-evolving template. If the blue fraction is
computed in such a way, then different values are found, even for a fixed galaxy
sample, because a non-evolving and an evolving template only match at $z=0$.
In fact,
Fairley et al. (2002) found that the blue fraction is higher if a bluer
rest--frame set of filters is used. Thus, some galaxies turn out
to be either blue or red depending on the selected filter set, although
the two classes should be separated.

\begin{figure*}
\psfig{figure=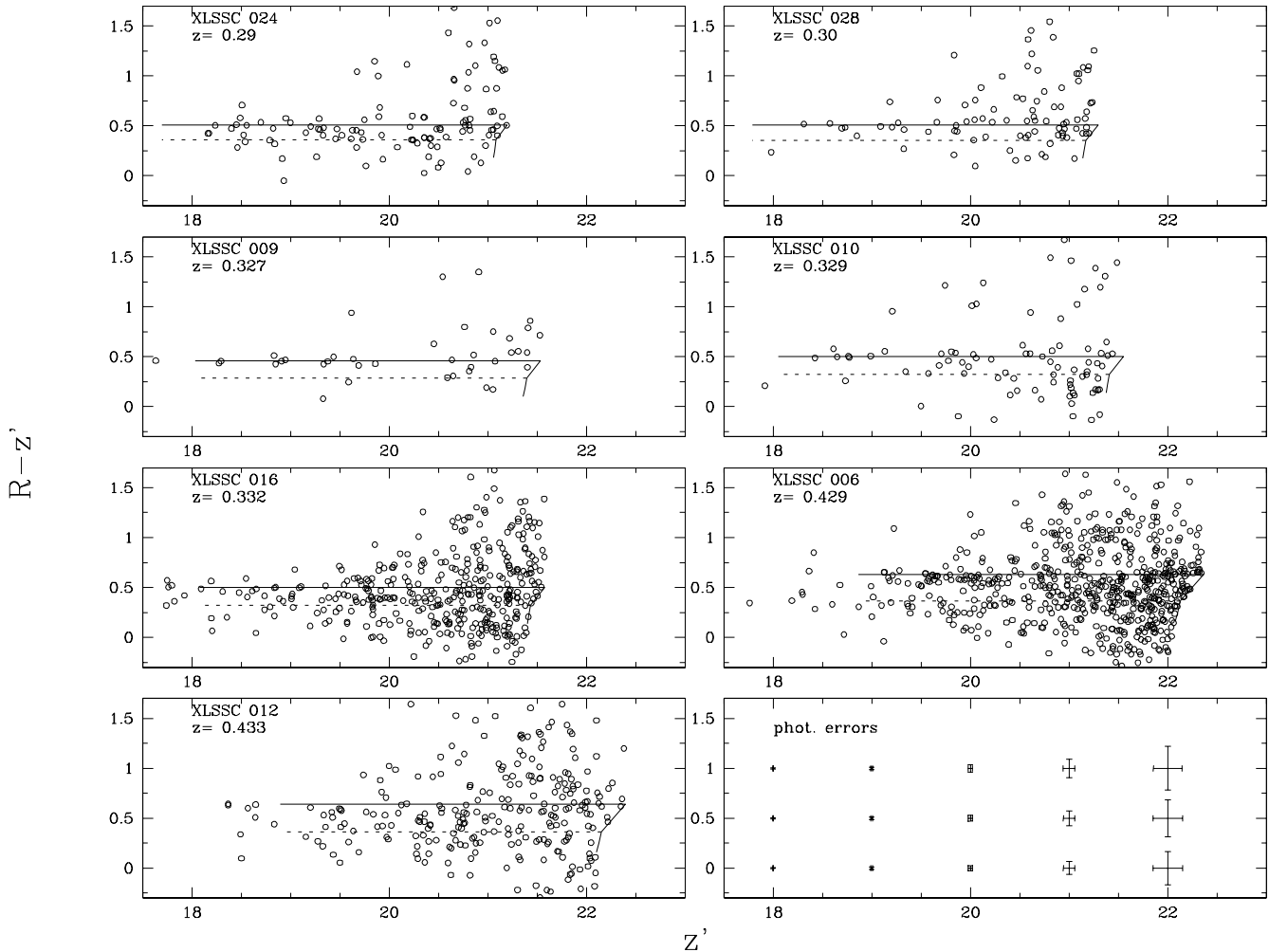,width=18truecm,clip=}
\caption[h]{Colour--magnitude diagram for galaxies within $r_{200}$. Only 
galaxies brighter than the evolved $M_V=-19.3$ mag (indicated with a spline
curve)
are shown. Colours are corrected for the colour--magnitude relation. 
The solid (dashed)
line marks the expected colour of an evolving E (Sa) spectral template. }
\end{figure*}

\begin{figure*}
\psfig{figure=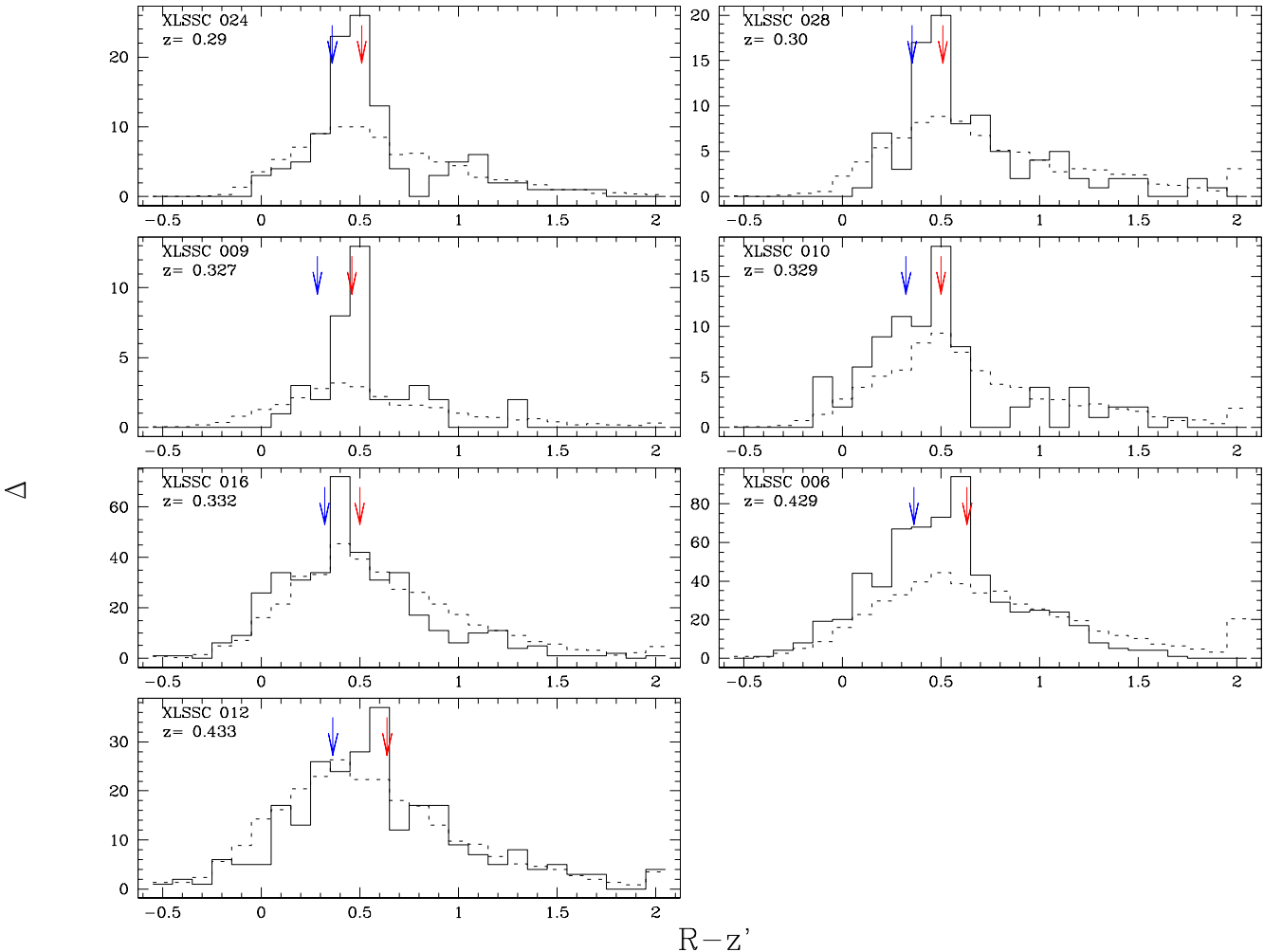,width=18truecm,clip=}
\caption[h]{Colour distribution of galaxies located within $r_{200}$ and
brighter than the evolved absolute magnitude $M_V=-19.3$ 
along the line-of-sight
of the cluster (solid histogram) and in the
control field (dashed histogram), normalized to the cluster area.
Colours are corrected for the colour--magnitude relation. 
The right (left) arrow
marks the expected colour of an evolving E (Sa) template. Colours are binned, and
consequently resolution is degraded, for display purposes only.}
\end{figure*}

The upper solid curve in the right panel of Fig. 1 reinforces the
conclusion of the
above discussion, but in the observer rest--frame. The continuous line
marks the expected $R-z'$ colour difference, in the observer bands, for an
evolving template having $\Delta (B-V)=0.2$ {\it today}, i.e. considering
our
evolving Sa spectral template.  The dashed curve illustrates the  $R-z'$
colour difference that one would incorrectly use if no stellar
evolution was allowed for.
It has
been computed for a non--evolving Sa template. Finally, the circles show
the $R-z'$ colour difference one should derive by using non--evolving
templates taken from Coleman, Wu \& Weedman (1980), as usually done. There 
is a rather good agreement between the latter track and our non--evolving Sa track
over a large redshift range ($0.3<z<0.7$): it reflects the fact that the spectra
of the two templates agree with each other at $z=0$ over a large wavelength
range and that our Sa model reasonably describes (at the
requested resolution) the observed spectra of Sa galaxies in the local 
Universe listed in Coleman, Wu \& Weedman (1980).

To conclude, we
definitively adopt an evolving Sa template to differentiate between
blue and red
galaxies, i.e. an evolving $\Delta$ colour cut as shown by the solid 
curve in the right panel of Fig. 1. Galaxies bluer than a Sa 
spectral--template
are referred to as ``blue", those redder, ``red". The blue fraction is
therefore computed with respect to a galaxy model that quietly forms
stars as our Sa model. Our choice has the advantage of focusing on
galaxy evolution, instead of focusing on observational problems
related to the filter choice or of assuming an unphysical universe, in
which the age of the Universe does depend on redshift,  but in which
the age of its content does not.

\section{Technical details}

Before proceeding with the calculation of the fraction of blue galaxies $f_b$, 
several additional operations need to be made: 

-- the colour red sequence is derived from
the median colour of the three brightest galaxies considered to be
viable cluster members, i.e. galaxies that are too blue or too bright
to be plausibly at the cluster redshift are discarded.

-- the slope of the observed colour--magnitude 
relation is removed from the data. The slope is an eyeball fit
to the observed colour--magnitude of
galaxies in the cluster center, in order to limit the background 
contribution.
We measure 0.025 colour mag per unit mag at the studied redshifts. 

-- the adopted radius that enclose an overdensity of 200 times the
critical density: $r_{200}$, computed from the relation

\begin{equation}
 r_{200} = { \sigma_{1D} \over  H_{0} \sqrt{ 30 [\Omega_m (1+z)^3 +\Omega_{\Lambda}]} } 
\end{equation}

(Mauduit, Mamon, \& Hill, 2005)
where $\sigma_{1D}$ is the cluster velocity dispersion. Found values
are listed in Table 1.

-- the center of the cluster is defined by the position of the brightest
cluster member (BCM), with one exception: XLSSC 006 has two BCMs,
and we took the cluster center at the middle of the two. The adopted
center is compatible with the detected X--ray center. Their precise
location is unimportant for measurements performed within $r_{200}$

-- Galaxies redder than an Sa are referred to as red galaxies (sect 4),  but
how far in the red direction should we integrate the colour distribution?
We adopted several cuts (including $+\infty$), and in six out of seven
cases, we find no evidence for a bias in the measured $f_b$ for any cut
redder than the colour of an E +0.05 mag, i.e. we find no statistical
evidence for a cluster population redder than the colour--magnitude
sequence plus 0.05 mag. Actually such a population is not 
expected from population synthesis models, because
the reddest model galaxies have the colour of the red sequence  galaxies.
By keeping the smallest value (the colour of an E+0.05 mag) we maximize the
S/N of the blue fraction determination, without biasing the measurement.

-- When selecting the background region, we choose the most representative
realization of the control field: all the regions which are not associated
with the target, i.e. such that $r> 2 r_{200}$. The precise radius used
(say $r/r_{200}> 2$ or $5$) is irrelevant, because the contribution of
galaxies in the cluster outskirts is negligible with respect to the number
of field galaxies in our huge control area (approximatively 0.3 deg$^2$).
Other researchers prefer instead to choose the background area in regions
particularly devoided of galaxies, hence unduly boosting the number of
members and apparently reducing the noise in the $f_b$ estimate.

-- We verified that our galaxy catalogs are complete down 
to $M_V=-19.3$ mag (and fainter magnitudes), as in previous works
(e.g. Andreon et al. 2004; Garilli, Maccagni, Andreon 1999).

\section{Results}

\subsection{Colour--magnitude and colour distribution}

The colour--magnitude relation and colour distribution of three (out of
seven) clusters in our sample are presented in Andreon et al. (2004a), and
discussed there with 15 additional clusters. Here we only want to discuss what
is directly relevant for the BO effect. 

Figure 2 shows the observed colour--magnitude relation for galaxies within
$r_{200}$ (including background galaxies), corrected for the
colour--magnitude slope (sec 5), and difference in seeing between the $R$
and $z'$ exposures (sec 2.1).  The solid line marks the expected colour of
the assumed  spectro-photometric E template discussed in sect. 4. There is
a good match between the expected and observed colours of the red sequence
for six out of seven cases. The red sequence of XLSSC 016 is slightly bluer
(by 0.05 mag) than expected, a feature that can be better appreciated in
Fig. 3. This single (out of seven), and admittedly small, offset is not in
disagreement with our error estimate for the colour calibration of about
$\la 0.03$ mag (Andreon et al. 2004a), and therefore such a minor mismatch
has been corrected for (by shifting the $R-z'$ colour by this amount), in
the $f_b$ determination, but has been left untouched in Figs. 2 and 3
to allow the reader to appreciate it. 
Unduly neglecting the above correction induces a bias (actually a
systematic error) of 0.01 in $f_b$. 
The error bar on $f_b$ (including everything in the error budget) 
turns out to be 16 times larger.

Figure 3 shows the colour histograms of galaxies brighter than the
evolved $M_V=-19.3$ mag located along
the line-of-sight of the cluster (solid histogram) and
in the control field (dashed histogram, $\sim 0.3$ deg$^2$), normalized
to the cluster area. The control field is taken from the same image 
where the cluster is observed, and hence shares the same
photometric zero--point
and quality. Therefore, any systematic photometric error largely simplifies in the
blue fraction determination, because both colour distributions are
shifted by the same amount (including the case of XLSSC 016).

\begin{table}
\caption{Blue fractions of individual clusters for galaxies within $r_{200}$}
\begin{tabular}{lrrll}
\hline
 Name   & $N_{gal}$ & error & $f_b$ & 68 \% c.i. \\
\hline
XLSSC 024 & 24 & 8 & 0.09 &  $[0.02,0.17]$  \\
XLSSC 028 & 14 & 7 & 0.06 &  $[0.01,0.11]$  \\
XLSSC 009 &  9 & 5 & 0.09 &  $[0.01,0.17]$  \\
XLSSC 010 & 24 & 8 & 0.51 &  $[0.33,0.68]$  \\
XLSSC 016 & 51 & 15 & 0.45 &  $[0.29,0.61]$  \\
XLSSC 006 & 204 & 21 & 0.43 &  $[0.38,0.48]$  \\
XLSSC 012 & 8 & 7 & 0.16 &  $[0.02,0.31]$    \\
\hline                                                   
\end{tabular}
\quad\break
$N_{gal}$ is the number of galaxies inside $r_{200}$ and brighter
than the evolved $M_V=-19.3$ mag. \hfill \break
\end{table}

\begin{figure*}
\psfig{figure=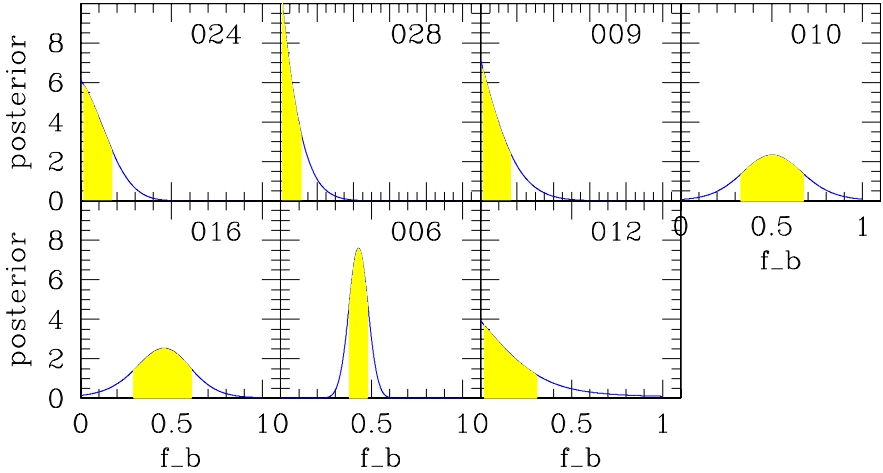,width=12truecm,clip=}
\caption[h]{Probability for $f_b$ at $r_{200}$ assuming
a uniform prior. The shaded regions delimit the 68 \% interval (error). At its
center lies our point estimate of the cluster blue fraction. Each panel
is marked by the last three digits of the cluster name.}
\end{figure*}

\begin{figure*}
\psfig{figure=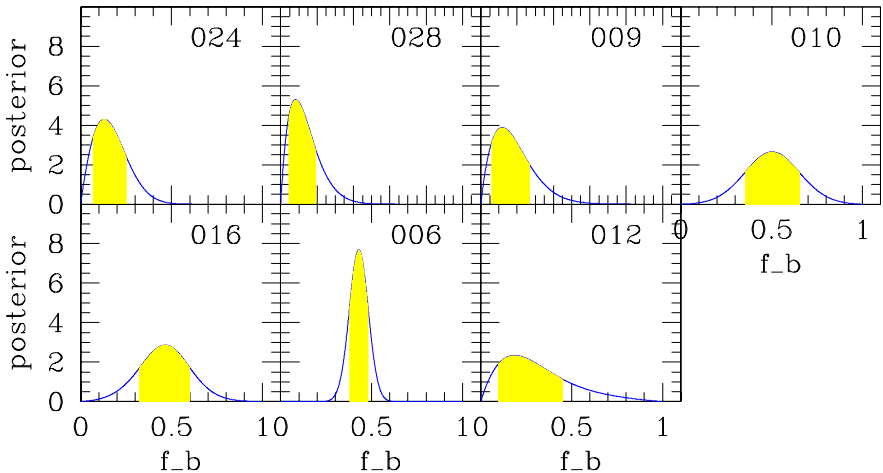,width=12truecm,clip=}
\caption[h]{As Fig 5, but for 
a parabolic prior.} 
\end{figure*}

\subsection{Blue fractions for individual clusters}

Table 2 summarizes our point estimate of 
the cluster richness, the blue fraction $f_b$, and 
its associated error, computed as
described in Appendix B and C. 
Shortly, we introduce methods of widespread use in the
statistical community, but largely unused in previous BO studies,
which are more robust than traditional methods. Instead of introducing
an estimator for the blue fraction and of providing
a point estimate of it which, in the long run (i.e. if
we were allowed to repeat the observations a large number of times), 
tends to the quantity aimed to measure (the blue fraction), we compute
the probability of each value of the blue fraction, given the data,
using the Bayes theorem of statistics. 
Bayesian inference is free from logical
contradictions of assigning negative (or complex) values to
positively defined quantities, that affected many previous
BO studies.

Richness ($N_{gal}$ in Table 2) is computed for galaxies
brighter than the evolved $-19.3$ mag and are located inside $r_{200}$.
Our clusters are quite poor, on average, although they show
a large range of richnesses.

Figure 4 shows the (posterior) probability that our
clusters have a fraction $f_b$ of blue galaxies within $r_{200}$ assuming 
a uniform prior. The 68 \% central credible intervals (errors) are drawn
as shades. They are usually small ($\sim \pm 0.1$), in spite of the fact
that many of our clusters contain few members. 
Figure 5 is similar to Fig 4, but under
a different assumption for the prior
(an upside-down parabola in the $[0,1]$ range and $0$ outside), 
in order to quantify
the robustness of the results on the assumed prior. The latter prior
quantifies the expectation of some readers, 
who believe that a Butcher--Oemler effect exists, i.e. who believe that
low values of the blue fraction are unlikely a priori. The parabolic
prior encodes such a belief, un--favouring low values of the blue fraction. 
Comparison
between Figs 4 and 5 shows that our point estimate for the cluster blue fraction
(the median, that by definition falls in the center
of the highlighted region) and its error (the width of the 
highlighted region) are only marginally affected by the choice of
the prior, if affectes it at all.

\begin{figure}
\psfig{figure=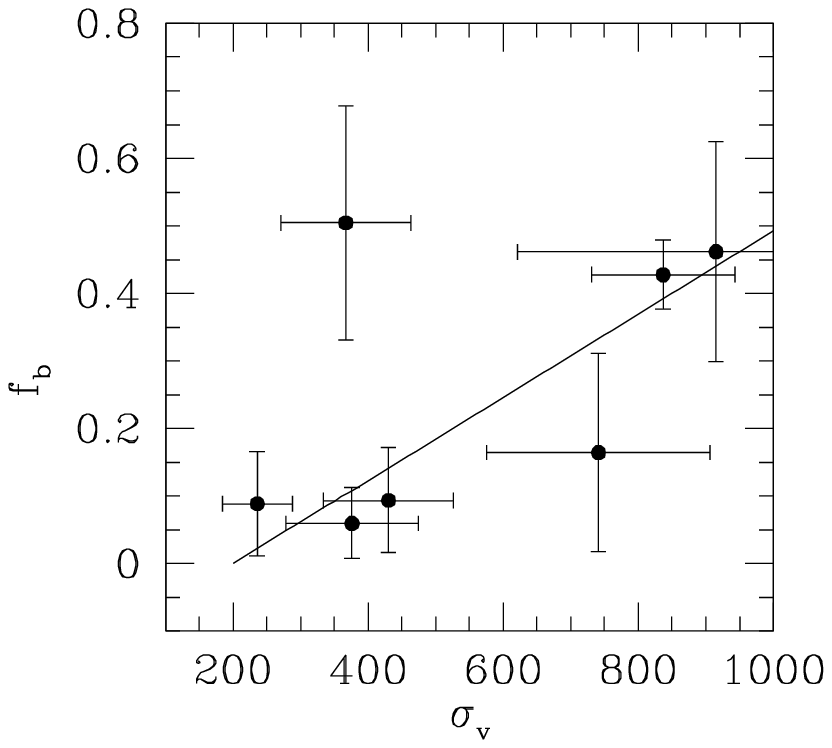,width=8truecm,clip=}
\psfig{figure=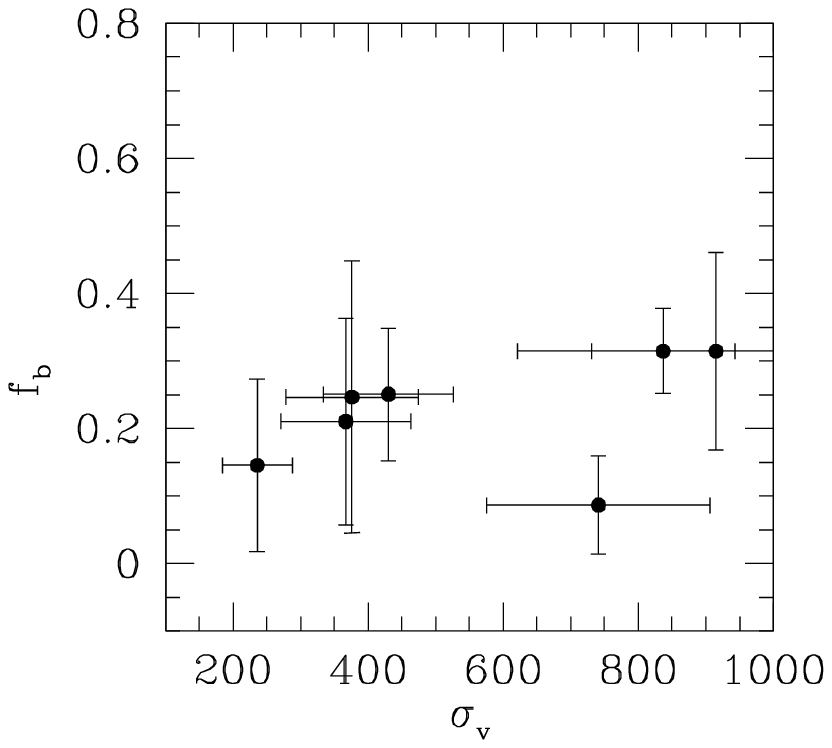,width=8truecm,clip=}
\caption[h]{Relationship between cluster velocity dispersion and
blue fraction within $r_{200}$ (top panel) and within $r_{200}/4$
(bottom panel).
In the top panel, the linear model favoured by the data is shown.
} 
\end{figure}

Three clusters have a blue fraction within $r_{200}$ of about 0.4,
whereas the other four clusters display a blue fraction of the order, or
less than, 0.1. More precisely, the richest clusters seem to possess  
the largest blue fractions.
What is the statistical significance of such a relationship, shown
in Fig 6?  Liddle (2004) reminded the astronomical community of
the difficult problem of model selection, i.e. in our case, 
to establish whether existing data support a
model in which the blue fraction $f_b$ depends on $\sigma_v$.
Our compared models (a constant $f_b$ vs a linear relationship between
$f_b$ and $\sigma_v$) are nested and regularity conditions hold in
our case.
The likelihood ratio turns out to be $2 \Delta \log \mathcal{L} \sim 6.6$ 
when adding
one more parameter. Therefore, under the null hypothesis (a constant
$f_b$) there is a 1 \% probability to observe a larger likelihood ratio by
adding one more parameter. 
Furthermore, the Bayesian Information Criterium (BIC) introduced by
Swartz (1978), and described in various statistical textbooks (and
also in Liddle 2004) offers another way to look at the same problem, in
the Bayesian framework. 
A value of 6 or more is regarded as strong
evidence against the model with a larger value of BIC 
whereas a value of two is
regarded as positive evidence (Jeffreys 1961). We find $\Delta BIC=5.8$ in
favour of the model $f_b \propto k (\sigma_v-200)$. To summarize,
there seems to be some good evidence for the existence of
a linear relationship between $f_b$ and $\sigma_v$.

However,
the adopted model appears to be unphysical, because for clusters having
$\sigma_v<200$ km s$^{-1}$, it predicts $f_b <0$. 
A more complex model is required, that perhaps
flattens off at low $\sigma_v$, avoiding unphysical $f_b$ values. At
this moment,
we consider such a model too complex, given the available set
of data. Evidence for a possible correlation is
recognized but it is considered far from being definitive.

Evidence for a correlation between the blue fraction and the velocity
dispersion largely disappears when choosing a smaller reference radius
(say $r_{200}/2$ or $r_{200}/4$), as shown in the bottom panel of Fig. 
6 for $r_{200}/4$.
Of course, a shallow relationship could be present, but our data do
not unambiguously favour it, because the relationship, if any, is swamped 
by the relative importance of
errors. The possible
lack of a relationship between the central blue fraction
and mass (measured by the cluster velocity dispersion) seems to confirm
a similar lack of correlation between the cluster X-ray luminosity (a
tracer of mass) and the central blue fraction (Andreon \& Ettori 1999; Fairley
et al. 2002).

At low redshift ($z<0.1$), Goto et al. (2003) and Goto (2005) tentatively
conclude from a larger sample of clusters that there is no evidence for a
relationship between the blue fraction and the cluster mass. However, their
definition of the blue fraction is different from ours, and their
statistical analysis is very different (for example Goto et al.
2003  have observed several clusters with unphysical values for the blue
fraction, see their Fig 1). Similarly, Balogh et al. (2004) find no
evidence at $z<0.08$  for a relationship between the fraction of blue galaxies
inside the virial radius and the velocity dispersion, although, admittedly,
fairly large uncertainties affect their results, 
besides another definition of what is  
``blue". Whether the relationship  sets itself at redshifts higher than
those probed by Goto et al. or Balogh et al., or whether
it is masked at low redshift
because of their various blue fraction definitions or because of
the way the analysis is performed, or, finally, is the result of a small
sample at $z\sim0.35$, is still  a matter to be investigated.

\begin{figure}
\psfig{figure=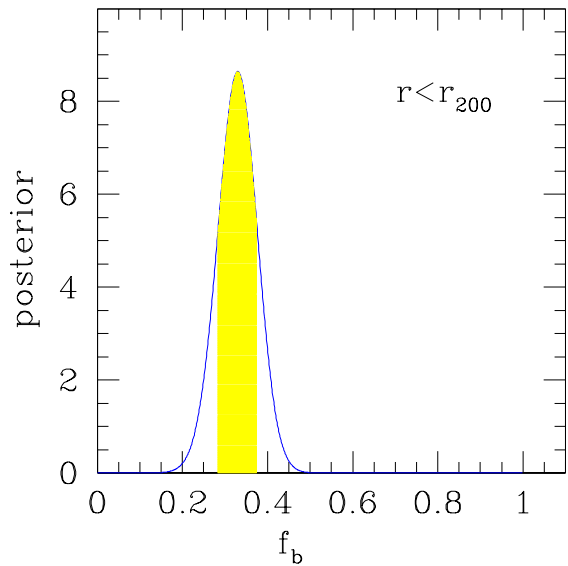,width=8truecm,clip=}
\caption[h]{Constraints on $f_b$ for the combined
sample.} 
\end{figure}

\subsection{Composite sample}

\subsubsection{Blue fraction of the composite sample}

Figure 7 shows the (posterior) probability that the combined
sample has a blue fraction $f_b$, computed using recursively the
Bayes theorem. It is
bell--shaped and narrow, that makes
the blue fraction in the composite sample well determined and
independent on prior: $f_b=0.33 \pm0.05$.
The combined sample is formed by about 320 cluster galaxies within
$r_{200}$. 

What does this result mean in the presence of a {\it possible} relationship
between the velocity dispersion and the blue fraction? 
The existence of measurements claimed to be incompatible does not 
constitute  an absolute obstacle when computing
a sample average in the Bayesian framework, provided that the studied
sample constitutes a representative one. 
It is in our everyday experience to
compute means of a population in which the elements differ much
more between each other
than the uncertainties affecting the individual measurements
(cf. the average post-doc salary, the average human weight or height, etc.).
These averages require the sample to be a representative one, 
otherwise the computed average would lack its predictive power.
Our sample is small, but constitutes a representative sample of 
clusters (sect 3).

\begin{table}
\caption{Radial dependence of the blue fraction of the combined sample}
\begin{tabular}{lrrll}
\hline
Sample  & $N_{gal}$ & error & $f_b$ & error  \\
\hline
$r<r_{200}$ & 321 & 32 & 0.33 & 0.04 \\
$r<r_{200}/4$ & 136 & 13 & 0.24 & 0.04 \\
$r_{200}/4<r<r_{200}/2$ & 109 & 16 & 0.30 & 0.07 \\
$r_{200}/2<r<r_{200}$ & 78 & 25 & 0.46 & 0.10 \\
$r_{200}<r<1.5 \ r_{200}$ & 48 & 25 & 0.55 & 0.14 \\
\hline   
field & 83 & 10 &  0.73 & 0.05 \\  
\hline                                              
\end{tabular}
\end{table}

\subsubsection{Radial dependence of the blue fraction in the composite sample}

Different physical mechanisms are thought to operate in different
environments (see Treu et al. 2003 for a summary) and thus, by identifying
where the colour of galaxies starts to change, we can hope to identify the
relative importance of such  mechanisms. For this reason, we studied the
radial dependence of the blue fraction $f_b$ as usually done in the
literature, by splitting the data in radial bins. We arbitrarily choose
$[0,1/4]$,$[1/4,1/2]$, $[1/2,1]$ and $[1,1.5]$ in units of $r_{200}$, for
simplicity.  In the outermost bin we were forced to drop XLSSC 016, because
$1.5 r_{200}$ lies farther away than the mid-distance between XLSSC 016 and the
nearest cluster to it, as seen projected on the plane of the sky, and,
therefore, this radial bin is potentially  contaminated by galaxies
belonging to the other cluster. Note that its inclusion, or exclusion, in
the other radial bins does not affect the derived values, and therefore our
conclusions. Table 3 lists the found values.

Figure 8 (solid points) shows that the blue fraction increases with the
clustercentric distance, from $0.24 \pm0.04$ in the innermost bin, to
$0.46 \pm 0.10$ and $0.55 \pm0.14$ in the two outermost bins:
galaxies at the center of clusters are found to have a
suppressed star formation (redder colours) compared to those at
larger clustercentric radii.

Have we reached the field value of the blue fraction?   Using the
spectrophotometry listed in COMBO-17 (Wolf et al. 2004),  that encompasses
$1/4$ deg$^2$ of the Chandra Deep Field region, we have selected the
galaxies brighter than the (same evolving) absolute magnitude limit adopted
in our work, and in the same redshift range ($0.29<z<0.44$).
There are 83 galaxies, of which 61 are bluer than an Sa evolving
template. We, therefore, infer a blue fraction of $0.73 \pm0.05$,
arbitrarily plotted at $r/r_{200}=2.5$ in Fig. 8. In the above calculation,
we were forced, for lack of information, to neglect redshift errors and
errors on the photometric corrections applied by the authors to compute
absolute magnitudes. 

\begin{figure}
\psfig{figure=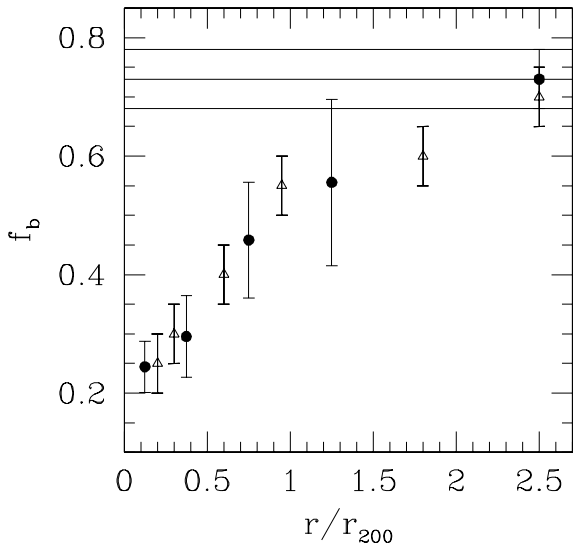,width=8truecm,clip=}
\caption[h]{Differential radial profiles.
Filled dots represent the blue fractions 
from our own photometry and analysis. 
Open
triangles correspond to the fraction of galaxies with normalized
star formation rates larger than 1 solar mass per year
from Lewis et al. (2002).
Our field value is 
arbitrarily set at $r/r_{200}=2.5$ for display purpose. 
} 
\end{figure}

The blue fraction is found to steadily increase from the cluster core 
to the field value.

The important point to note in Fig 8 is that the influence of the cluster 
reaches large radii. There are two possible explanations for the above
result. First, the mechanism affecting the galaxy colours reaches
large radii. In such a case, ram pressure stripping, tidal halo
stripping and tidal triggering star formation (just to mention a few, 
see e.g. Treu et al. 2003
for definitions and other examples) are ruled out as 
{\it direct} causes of the observed trend, because
all of them are short-range mechanisms being effective in the cluster
center only.

Alternatively, the trend might be produced by a significant population of
backsplash galaxies, i.e. satellite galaxies that once were inside the
virial radius but now reside beyond it,  as suggested on theoretical
grounds by Gill et al. (2005). These authors claim that about the same
number of infalling galaxies and backsplash galaxies should be at $r \sim
r_{200}$. Under the reasonable assumption that  infalling galaxies have a
blue fraction equal to the field one, and rebounded galaxies have a blue
fraction equal to the central one, the expected blue fraction $f_b$ at
$r_{200}$ should be about 0.49  ($=(0.73+0.25)/2$), in good agreement with
the observed value, given support to the backsplash population alternative,
in agreement with models and observations presented in Balogh, Navarro  \&
Morris (2000). From a strict statistical point of view, this possibility is
favoured because it provides a sharp prediction verified by the
observations. The kinematical predictions of Gill et al. (2005) are also
in qualitative agreement with observations of the Coma cluster: blue
spirals (identified as the infalling population) have a higher velocity,
relative to the cluster center,  than red spirals (identified as rebounded
objects) and early--type populations (Andreon 1996).

The possible existence of a backsplash population, whose importance seems
hard to quantify from theoretical grounds, requires to keep in stand-by our
conclusion, as well other conclusions based on the (often implicit)
hypothesis that the population observed at large radii is uncontaminated by
rebounded galaxies (e.g. McIntosh, Rix, \& Caldwell, 2004). For the very
same reason, one should keep in hold the interpretation of the
morphology--density (or whatever density--dependent trends in  population
properties, such as the strong emitter fraction),  because it could either be
the result of mechanisms operating at the studied density, but
also  the result of different degrees of contamination (at different 
distances from the cluster center) by the  backsplash population. We are
not questioning the existence of the segregation, but the way one may
interpret it.

\section{Discussion \& Conclusions}

\subsection{Comparison with previous works}

Comparison of our results with other ones requires to pay attention
to the prescriptions adopted to define the blue fraction,
to the way the cluster sample is built, and, sometime, to
the adopted statistical approach.

\subsubsection{Evolution of the blue fraction}

The most similar work to ours is the seminal Butcher \& Oemler paper, from
which we modelled our prescriptions. By selecting a small subsample of
clusters of richness similar to ours but located in the nearby universe
($z\sim0.02$, where BO and our prescriptions are identical) they find $f_b$
values within $r_{30}$, the radius that includes 30 per cent of cluster
galaxies, in the $0.02$ to $0.19$ range, with a typical error of
$\pm0.03$. This range of values is not significantly lower, considering the
various sources of uncertainties, than our central value $f_b=0.24\pm0.04$, to
claim that the two values are different at a high significance level,
especially taking into account the fact that the Butcher \& Oemler errors are
sometimes optimistically estimated (Andreon, Lobo \& Iovino 2004).

de Propris et al.'s (2004) large sample of nearby clusters matches our
sample in terms of richness: we find for their sample\footnote{We thank R. de
Propris for giving us their blue fraction within $r_{200}/2$} an average
$N_{gal}$ of 30 galaxies and a blue fraction inside $r_{200}/2$ of $0.17$,
taking into account that the blue fraction is a binomial deviate (the authors
assumed it to be a Gaussian and find $f_b=0.13$). The error due to the sample
size is negligible (0.01) because their sample is large. However, the largest
source of uncertainty in their work comes from their large photometric errors
(they use photographic plates). Such photometric errors induce a bias in the
blue fraction that, as discussed in Sect 3.1, is difficult to correct for
(Jeffreys 1938, Eddington 1940), and is neglected by the authors.  After
accounting for minor differences in the luminosity cuts between de Propris et
al. and BO and for the mentioned Malmquist bias, the estimated blue fraction
within $r_{200}/2$ in the de
Propris et al.'s (2004) sample becomes $\approx 0.25$, but with an error hard
to quantify. Inside $r_{200}/2$ we find  $f_b=0.26\pm0.04$, which identical to
what found in the de Propris et al.'s large nearby sample.

To summarize, our $z\sim0.35$ sample matches in terms of richness the nearby
samples in Butcher \& Oemler (1984) and, especially, de Propris et al. (2004) 
and shows equal blue fractions within $r_{30}$ and $r_{200}/2$, i.e. no
Butcher-Oemler effect is seen. It is to be
noted that our sample has an almost identical size and redshift distribution as
the high redshift clusters in the BO sample, and thus that our lack of
detection of a BO effect is not due to a smaller or closer sample.

The compared clusters matches in terms of richness, but are constructed using
different selection criteria, because the low redshift sample is an optically
selected one, while our cluster sample is x-ray selected. As mention in sec. 1,
our x-ray selection is chosen to minimize the observational bias on $f_b$, and
hence to derive a fair measure of the blue fraction. At low redshift, we are
not aware of any reason why $f_b$ should be biased {\it at a fixed richness}
for an optically selected sample, such as the ones of Butcher \& Oemler (1984)
and de Propris  et al. (2004): why clusters of a given richness and rich  in
blue galaxies should be over/underrapresented in cluster catalogs of the nearby
universe? Therefore, even if the selection criteria used to build the compared
cluster samples are different, the comparison of the blue fractions is safe,
because both cluster samples provide unbiased values of $f_b$.

There are hints that confirm the constancy of the blue fraction at even larger
redshifts. ALI04 show evidence for a {\it low} blue fraction at $z\sim0.7$.
Recently, Tran et al. (2005) also find a {\it low} blue fraction ($f_b=0.13$) 
for a cluster at $z\sim0.6$, computed inside a cluster portion that, if not
rigorously identical to the one prescribed by Butcher--Oemler, does support the
{\it non} existence of a Butcher--Oemler effect.  Both works adopted a
non-evolving $\Delta$. If an evolving $\Delta$ is used, the derived blue
fraction at high redshift would even be lower than claimed, giving further
support to our conclusion. ALI04 have also disproved all the reported
literature evidence accumulated thus far for the existence of a 
Butcher--Oemler effect, i.e for a change of the blue fraction inside $r_{30}$. 

All the above suggests that the fraction of blue galaxies, computed
by separating the galaxies using a population formed by stars whose age 
increases at the same rate as the universe age increases, does not evolve.
Or, if the reader prefers, there is no systematic drift from the blue
to the red classes of galaxies as the look--back time evolves, 
between $z\sim0$ and $z=0.44$.

\medskip

A result similar to the one depicted in Fig. 8 is presented in Lewis et al.
(2002) based on nearby clusters. Our results are in qualitative agreement
with theirs since we also find that the cluster affects the fraction of active
galaxies up to the virial radius. Lewis et al. (2002) have studied a {\it
nearby} cluster sample composed of 440 member galaxies inside the virial
radius (vs our sample of 320). Figure 8 shows that their fraction of galaxies
with star formation rates, normalized to $M^*$, larger than 1 solar mass per
year (triangles) nicely compares with our derived fraction of blue galaxies.
Our error bars for their points show the expected central 68 per cent credible
intervals, only accounting for sampling errors, computed by us from a
straightforward application of statistics. In Lewis et al. (2002), the sample
is split in classes very similar to ours and BO: in fact, our spectro-photometric Sa
has a star formation rate, normalized to $M^*$, equal to their adopted
threshold (one solar mass per year per $M^*$ galaxy) if $M^*=8.2 \ 10^{10}
M_{\odot}$, a value well inside the range of values usually observed (e.g.
Blanton et al. 2001; Norberg et al. 2002). I.e. what is called blue
by them is also called blue by us, on average. It is not surprising,
therefore that integrating the Lewis et al.'s (2002) blue profile within
$r_{200}/2$ gives a blue fraction identical to the one observed
in the Propris  et al.'s (2004) sample ($0.26$ vs $0.25$), further
supporting the similarity of the two classes (blue by colour and blue by
star formation rate).

Their cluster sample has an
overlapping, but different, range of masses (velocity dispersion) with respect
to our sample: our richest clusters have a velocity dispersion typical of the
average values of Lewis et al. (2002) clusters. However, their profile is
only marginally affected, if at all, by separating clusters in (two) velocity
dispersion classes (Lewis et al. 2002). Furthermore, Gomez et al. (2003)
indirectly confirm that the radial profile is not too much affected by
differences in cluster mass, by studying a sample of nearby clusters having
velocity dispersions similar to our sample. Therefore, differences
in the way cluster sample are built seems not to affect the derived
``blue" profile.

The agreement between Lewis et al. (2002) and our blue fraction profiles is
almost perfect; however the however the studied clusters are located at quite different
look back times:  clusters in Lewis et al. are in the very nearby universe (at
$z\sim0.07$), whereas our clusters have $z\sim0.35$, implying a $\sim3$ Gyr
time difference for the adopted cosmology. As long as the
separation of galaxies in classes by Lewis et al. (2003) and in our work
is similar, the agreement of the two radial profiles
means that there is no evolution of the blue fraction between $z\sim0$ and
$z\sim0.35$, from the cluster center to the field value.

\subsubsection{Disagreements or different ways in interpreting the data?}

Considering a much more luminous X-ray (and therefore massive) sample of
clusters at intermediate redshift, Fairley et al. (2002) find an increasing blue
fraction as a function of the clustercentric distance, up to 2 $r_{30}$, in good
agreement with the results we found over much larger clustercentric distances. A
quantitative comparison between the two pieces of work is however impossible,
because there are too many uncontrolled variables that are allowed to change
between these.  The authors find a blue fraction $f_b\approx0.2 \pm0.1$, in
agreement with the value we observe in the cluster center $f_b=0.24\pm0.04$.
However, we believe that this apparent agreement  largely arises by chance.
First, the authors used a non-evolving template to separate the galaxies in red
and blue classes, and find two sets of (different) values for their two sets of
available colours. Secondly, they considered higher redshift than we do, by
observing clusters with comparable exposure times but with smaller telescopes
(2.5m vs 4.0m). In spite of their expected larger errors, they neglect the
effect of photometric errors on their blue fraction estimates (sect 3.2). Third,
they do not adopt an evolving luminosity limit. And, finally, a comparison of
the values derived in the two works requires an extrapolation, because clusters
with very different masses (X--ray luminosities) are considered.

Ellingson et al. (2001) performed a study quite different from ours, and
adopt a galaxy separation that is the same irrespective
of redshift (i.e. of galaxy age), because they decomposed their
spectra on non-evolving spectral templates. 
Their claim for a change in the population
gradient is just a restatement of the fact that the blue fraction is higher
everywhere in the cluster and in the field because galaxies are bluer when they
were younger, i.e. is not informative about processes running in clusters or in
the field, but just informative about aging. These authors would observe an
evolution of the gradient even if galaxies would be kept isolated from the
surrounding environment and the infall in the cluster would be fixed (i.e. no new
galaxy falls in the cluster, and galaxies are kept fixed at their observed
position): their ``old population" fraction increases going from high to low
redshift  because galaxies become older, and the effect is more marked  at large
clustercentric radii than in the centre,  because in the cluster core the  ``old
population" fraction is already near to 1 and cannot take values larger than 1.
We, instead, choose to reduce by one the number of parameters, removing the age
dependency by using an evolving (Sa) template.

In summary, the referenced analysis do not reveal results in disagreement with
our own work, although their interpretation may sometimes be different (or even
opposite to ours).

\subsection{Conclusions}

This paper revises the definition used to separate galaxies in two colour
classes in a way that takes in to account the reduced age of the Universe at
higher redshift.  It is nowadays uninteresting to know whether the fraction of
blue  galaxies changes with redshift in a way that is different from the
expectation of a model that we now know is unphysical (that has the same age at
all redshifts). If the model is unphysical, there is no need to make
observations to rule it out. A stellar population whose age does not change in
a Universe whose age instead changes, as it is supposed by using a non-evolving spectral
template  (or a fixed $\Delta$, i.e. the BO prescription), is clearly
non-physical. It was useful a long time ago to show that a universe with the
same age at all redshifts is rejected by observations. However, nowadays we can
attack a more essential question: 
to know whether galaxies evolve differently from a reference evolution that is
physically acceptable. Our measurements of evolution are, therefore,
zero-pointed on the evolution
of an object whose age increases as required by the
current cosmological model. We select a  spectro-photometric Sa to conform to
the BO prescription in the local universe.  Effectively, this is
a change in
perspective: we should no longer attempt to reject an unphysical
universe, in which the age of the Universe does depend on redshift, whereas the
age of its content does not, but we should study whether the observed
differences between the low and high redshift content are in agreement with
differences of the Universe age at the considered redshifts. 

Furthermore, we have introduced in our specific domain the tools of Bayesian 
inference (see Appendix), dramatically improving on previous approaches that
led some authors to claim that they have observed unphysical values
(such as blue fractions outside the $[0,1]$
range or negative star formation rates). Such tools allow us to use 
all our data without rejecting blue fractions measured at
large clustercentric radii, where the signal to noise is low,
contrary to previous researchers obliged to discard such data
(or claiming that they have observed unphysical values).

The main result of this work is that
{\it we find
that the cluster affects the properties of the galaxies up to two
virial radii at $z\sim0.35$}. 

We have measured the blue fraction of a {\it representative} sample of
clusters at intermediate redshift. Indeed, our sample is a random sampling of
a volume complete X-ray selected cluster sample. The X-ray selection has no 
cause-effect relationship on the cluster blue fraction, 
all the remaining parameters being kept fixed, to the best of our knowledge, and,
therefore, the studied sample consists of an unbiased one (from the blue
fraction point of view). Our statement should
not be over-interpreted, however, because we are only sampling a portion
of the X-ray parameter space: very X-ray luminous clusters are missing
in our sample because they are intrinsically rare, and
clusters with fainter X-ray emission than
the limiting flux are missing because they lie outside the sampled space.

Studied clusters show a variety of values for
the blue fraction, when the fraction is measured within $r_{200}$. The variety is
too large to solely be accounted for by errors. At smaller radii, instead,
the blue fractions are more homogeneous.
Actually, there is some evidence that the blue fraction within
$r_{200}$ increases with the cluster velocity dispersion, i.e. 
with the cluster mass,
whereas the increase at smaller radii is much smaller, if present at all.
Therefore, intermediate redshift clusters with the
largest masses show the largest fractions of star-forming galaxies, when
measured
within $r_{200}$. However, the evidence is good but not 
definitely conclusive
and still requires an independent confirmation.

The radial dependence of the blue fraction is quite shallow: it
smoothly and monotonically increases from the centre to the field. 
The latter has been determined according to our prescriptions 
using COMBO-17 data. 

The radial dependence (i.e. the blue fraction at every computed
clustercentric radius) is equal to the one recently found 
in a comparable sample of clusters, but in a 3 Gyr older universe, i.e.
at $z\sim0$ (Lewis et al. 2002). The agreement between the two 
derived profiles (amplitude and shape),
our blue fraction within $r_{30}$ and $r_{200}/2$ and the local similar determinations
(Butcher \& Oemler 1984, de Propris et al. 2004), the low blue fractions
at high redshift (ALI04, Tran et al. 2005),
all of these suggest that there is no colour evolution beyond 
the one needed to account for the different age
of the Universe and of its content. The above is found to hold
from the cluster core to the field value. Previous controversial
evidence from the literature 
assumed that the universe becomes older while
its content does not, and
overstated the significance of the evidence or compared
heterogeneously measured blue fractions (as shown in
ALI04).

The interpretation of the observed radial trend is complicated by 
the possible existence of a backsplash population. If the backsplash
population represents a negligible fraction of galaxies at a given 
clustercentric radius, then the large clustercentric distance at which
the cluster still produces some effect
rules out short--range scale mechanisms. However, the predicted backsplash 
population is precisely what is needed to explain our observed fraction
at $r_{200}$, given the fraction at the cluster center and in the field,
and also qualitatively accounts for different kinematics of galaxies
having different star formation rates (blue and red spirals) in the
Coma cluster.
If this is the case, mechanisms efficient in the cluster center only come
into play, because galaxies are affected when they reach the cluster
core, and are then scattered at large clustercentric radii where they
spend a lot of time and are observed. 

The possible existence of the backsplash population does not
offer us the possibility to draw a final inference about the
nature and the time scale of the processes that shape
galaxy properties in clusters. 
The backsplash mechanism is a physical one: it affects 
the interpretation of measured radial (or
density) trends drawn by us and
other authors, and forces us to keep in hold their
interpretations.
However, the infall pattern  turns out not to have changed during the last
3 Gyr, as measured by the identical blue fraction profiles at
$z\sim0$ and $z\sim0.35$, in spite of apparently contradictory
previous claims, based on the use of a fraction definition that 
has one more (uncontrolled) dependence.

\section*{Acknowledgments}

We warmly thank Giulio D'Agostini, for his advices on the
inference theory, allowing us to perform the 
$f_b$ measurements presented in this paper, as well as 
Andrew Jaffe, Andrew Liddle and Mike West for useful discussions
about Bayesian statistics and model selection.
Jon Willis is thanked for
his efforts in acquiring part of the CTIO data used in this paper 
and for useful comments.
Malcom Bremer, Pierre-Alain Duc and Ivan Valtchanov
are thanked for their comments on
an early version of this paper, and Gianluca Lentini for 
useful English hints.
GG acknowledges FONDECYT grant \#1040359 and
HQ acknowledges partial support from the FONDAP Centro de Astrofisica (CONICYT).
Part of this work was also performed in the framework of the IUAP P5/36 project, 
supported by the DWTC/SSTC Belgian Federal Service.
This paper is based on
observations obtained with XMM, with ESO (prop. 70.A-0733), 
Cerro-Tololo Interamerican Observatory (prog. 2000-0295, 2001-0316 and 
2002-0247) and Magellan telescopes at Las Campanas Observatory.

\appendix

\section{Statistical Inference: Velocity dispersion}

Velocity dispersions and their uncertainties  are computed
according to statistical inference textbooks in a Bayesian
framework, from the observed values of the galaxy redshifts, while
accounting for measurement  errors. We first numerically derive,
using a Monte Carlo simulation, the likelihood of observing $\sigma_v$
computed by using the scale
parameter introduced by Beers et al. (1990), given the observed
redshifts and redshift errors. Then, using
the Bayes theorem and adopting a uniform prior we derive the
probability that the cluster has a velocity dispersion $\sigma_v$,
given the observed values of redshifts. The posterior, for the
chosen prior, turns out to be very well described by a Gaussian.

Velocity dispersions (point estimates) and
uncertainties (68 per cent central credible intervals) are quoted in Table 1
and are robust to changes of priors: adopting a widely
different prior ($1/\sigma^2$), our point estimate of the cluster
velocity dispersion changes by 2 to 5 per cent of its uncertainty.

Derived velocity dispersions are corrected for the $(1+z)$ effect.

Our velocity dispersions have the properties to be non negative, and
their uncertainties do not include unphysical (negative or
complex) values of the velocity dispersion. While the above properties 
seem useless to state, it should be noted that they are non trivial
properties, because some unphysical velocity dispersions are still published.

It should be noted that, for the velocity dispersions presented here, the frequentist and Bayesian
derivation of the value of the velocity dispersion turn out to be
quite similar.

\section{Statistical inference: the cluster richness }

The cluster richness within $r_{200}$ is {\it not} naively 
derived using the common background subtraction
(e.g. Zwicky 1957, Oemler 1974):

\begin{equation}
n(clus)=n(total, cluster+field)-n(total,field)
\end{equation}

with obvious meanings for the symbols, using the {\it observed}
number of galaxies, because it potentially leads to
negative numbers of cluster galaxies, which is acceptable
for the estimator described above, but not for the true
value of the physical quantity aimed to be measured (the cluster richness).
Furthermore, frequentist confidence intervals may have
whatever size, including being empty or of vanishing length
(as actually occurs
precisely for the above expression when the right-hand side of eq. B1
is negative, e.g. 
Kraft et al. 1991).
Measurements derived from eq. B1, and their
confidence intervals, have not the properties we would like
richness and errors to have (for example, richness to be positive,
and errors to be be large when the uncertainty is
large, and to become small when the uncertainty on nuisance parameters 
decreases, etc.). These are well
known and discussed with several degrees of approximation in both
the frequentist and bayesian frameworks (Helene 1983; Kraft et al.
1991; Loredo 1992; Prosper 1998; D'Agostini 2003).

Results derived from Eq B1, when $n(total,
cluster+field) \approx n(total,field)$ are difficult to be
understood and used (say in computing averages, or when we
need to propagate the uncertainty from $n(clus)$ on a derived
quantity). This situation occurs 
for one of our clusters (XLSCC 012, if eq. B1 is used, but
we do not use it): its richness is $-2$ and its confidence
interval (at whatever confidence level) is empty
(Kraft et al. 1991).

``If the results
are to  be supposed to have any relevance beyond the original data''
(Jeffreys 1938), we believe that it is preferable
to quote the point estimate of the cluster
richness, given the data in hand, in place of the algebraic result
of eq B1. We compute the (posterior)
probability that the cluster has $n$ galaxies, and, when
needed, we summarize it quoting, as we do
for the velocity dispersion (appendix A) and the blue fraction
(Appendix C), the median and the 68 per cent central interval. 
Specifically, we assume a uniform prior, taking advantage of
the fact that the problem is mathematically
worked out by Kraft et al. (1991). We checked that an almost
identical result is obtained using a Jeffrey prior (the problem is
worked out by Prosper 1998), once differences in the type of credible
intervals are accounted for, i.e. that the result found is only marginally
affected, if at all, by the prior choice.

It is comforting to find that XLSSC 012, which has at least 12
spectroscopic confirmed members (Table 1), has some
hot emitting gas, and hence {\it does} exist and
has, as all clusters of galaxies, a positive number of galaxies, 
has a listed richness of 8 galaxies within $r_{200}$
(brighter than an evolved $M_V=-19.3$ mag), even if the (naive, but
widespread) application of eq. B1 attributes to it a negative number of
galaxies ($-2$) and an empty confidence interval.

Eq. B1 is routinely used in computing cluster luminosity functions
in presence of a background, starting with Zwicky (1957), Oemler (1974).
Andreon, Punzi \& Grado (2005) update their use.

We conclude this section by reminding that, both in the frequentist
and bayesian paradigms, the background subtraction (marginalization) 
does not require that the background in the cluster line of sight 
is equal to the {\it average} value
or equal to the one observed in the control field, but
only that it is drawn from the same parental distribution,
contrary to some astronomical misconceptions.

\section{Statistical inference: the blue fraction}

Many
blue fractions published in astronomical papers can be dramatically improved: 
although, by definition, the blue fraction is hardly bounded in the
[0,1] range (otherwise part of a sample is larger than the whole sample), 
it is often claimed that the observed value of the blue
fraction is outside the [0,1] range (data points outside this range
are present  in several BO-like papers).
In presence of a background, unphysical values frequently occur. 
The reason is that the blue fraction is
computed from:

\begin{equation}
f^o_b(clus)={n(blue, cluster+field)-n(blue,field) \over n(total,
cluster+field)-n(total,field)}
\end{equation}

with obvious meanings for the symbols. The unavoidable use of the
{\it observed} number of galaxies, instead of the (unknown) true ones, in
the above formula makes the result difficult to be understood, for
the very same reasons already discussed for the richness estimator
(eq. B1).
The use of the {\it observed} number of galaxies in
Eq. C1 allows to find negative values for the blue fraction (i.e. we would
claim that there are more red galaxies than galaxies of all colours)
or blue fractions larger than one (i.e. we would claim that there
are more blue galaxies than galaxies of all colours), statements that
are hard to defend\footnote{Frequentist statisticians know how 
to defend unphysical values and confidence intervals which 
contain unphysical values, but most astronomers probably
will have some problems in understanding what actually the
numbers provided by the frequentist paradigm mean, and will find
hard to use them, for example for computing a mean
over an ensemble.}.  This mainly occurs when Poissonian fluctuations
make background counts larger than counts in the cluster line of
sight, or when there are more blue galaxies in the background 
than in the cluster line
of sight.

Eq C1, adopted in Postman et al. (2005), forced these authors to
discard two of three of their $z>1$ clusters in the determination
of the spiral fraction.
Bayesian inference allows not to discard data, to derive
estimates that never take unphysical values, and (credible)
intervals that have the properties we would like errors to have.
As for the cluster velocity dispersion 
(appendix A) and richness (appendix B), 
we compute the (posterior) probability that the cluster has a
blue fraction $f_b$,
given the observed number of galaxies in the cluster direction (total
\& blue) and the expected number of background galaxies (total \&
blue) measured over a large control field.  All the
mathematical aspects of the above computation have been worked
out by D'Agostini (2004), who provides all requested details and the
exact analytic expression for the likelihood, to be used to derive the
posterior, given  our data. Such a posterior may be summarized by a few
numbers: the median (our point estimate of the cluster blue fraction)
and the 68 per cent central credible interval (our estimate of the
uncertainty), exactly as we did for the case of velocity dispersion 
and richness.

A trivial application of the Bayes theorem allows 
to account for errors on $r_{200}$ (due to the uncertainty on
$\sigma_v$).
We verified that, properly accounting for $r_{200}$ errors, our results
got unchanged, mainly because the blue fraction is a smooth and
slowly varying function of $r_{200}$.

It is interesting to note that a reasonable constraint on $f_b$ is achieved even
for XLSSC 012, in spite of the naive expectation that,  the cluster being poor
(eq. B1 would get $-2$ galaxies), and  the cluster richness appearing at the
denominator of eq C1, the error on the fraction is huge, and therefore the
corresponding determined blue fraction is of low quality. The correct inference
takes, instead, a different approach and quantifies what is qualitatively apparent
in Figure 3: to the left of the blue (left) arrow there is no evidence for an
excess of blue galaxies in the cluster line of sight. Under such a condition, how is
it possible that the cluster blue fraction gets large if almost no blue galaxy 
overdensity is observed?  Given that almost no cluster blue galaxies are there,
the cluster blue fraction is low, and the number of red galaxies sets how rich
the cluster is and therefore   ``how low" the fraction is.

\bsp

\label{lastpage}

\end{document}